%% file: ms.tex
\documentclass[iop]{emulateapj}

\usepackage{natbib}
\usepackage{amsmath}
\usepackage{url}

\begin{document}

\title{Half of the Most Luminous Quasars May Be Obscured:
  Investigating the Nature of WISE-Selected Hot, Dust-Obscured
  Galaxies}

\author{R.J.~Assef\altaffilmark{1}, 
  P.R.M.~Eisenhardt\altaffilmark{2},
  D.~Stern\altaffilmark{2},
  C.-W.~Tsai\altaffilmark{2,3},
  J.~Wu\altaffilmark{4},
  D.~Wylezalek\altaffilmark{5},
  A.W.~Blain\altaffilmark{6},
  C.R.~Bridge\altaffilmark{7}
  E.~Donoso\altaffilmark{8},
  A.~Gonzales\altaffilmark{7,9,10},
  R.L.~Griffith\altaffilmark{11},
  T.H.~Jarrett\altaffilmark{12}}

\altaffiltext{1}{N\'ucleo de Astronom\'ia de la Facultad de
  Ingenier\'ia, Universidad Diego Portales, Av. Ej\'ercito Libertador
  441, Santiago, Chile. E-mail:~{\tt{roberto.assef@mail.udp.cl}}}

\altaffiltext{2}{Jet Propulsion Laboratory, California Institute of
  Technology, 4800 Oak Grove Drive, Pasadena, CA 91109, USA.}
    
\altaffiltext{3}{NASA Postdoctoral Program Fellow}

\altaffiltext{4}{UCLA Astronomy, PO Box 951547, Los Angeles, CA
  90095-1547, USA}

\altaffiltext{5}{European Southern Observatory,
  Karl-Schwarzschildstr.2, D-85748 Garching bei M\"unchen, Germany}

\altaffiltext{6}{Physics \& Astronomy, University of Leicester, 1
  University Road, Leicester LE1 7RH, UK}

\altaffiltext{7}{Division of Physics, Math, and Astronomy, California
  Institute of Technology, Pasadena, CA 91125, USA}

\altaffiltext{8}{Instituto de Ciencias Astron\'omicas, de la Tierra, y
  del Espacio (ICATE), 5400, San Juan, Argentina}

\altaffiltext{9}{Department of Earth, Atmospheric and Planetary
  Sciences, Massachusetts Institute of Technology, Cambridge, MA
  02139, USA}

\altaffiltext{10}{Scripps College, 1030 Columbia Avenue, Claremont, CA
  91711, USA}

\altaffiltext{11}{Department of Astronomy and Astrophysics, The
  Pennsylvania State University, 525 Davey Lab, University Park, PA
  16802, USA}

\altaffiltext{12}{Astrophysics, Cosmology and Gravity Centre (ACGC),
  Astronomy Department, University of Cape Town, Private Bag X3,
  Rondebosch 7701, South Africa}

\begin{abstract}
The WISE mission has unveiled a rare population of high-redshift
($z=1-4.6$), dusty, hyper-luminous galaxies, with infrared
luminosities $L_{\rm IR} > 10^{13}~L_{\odot}$, and sometimes exceeding
$10^{14}~L_{\odot}$. Previous work has shown that their dust
temperatures and overall far-IR spectral energy distributions (SEDs)
are significantly hotter than expected to be powered by
star-formation. We present here an analysis of the rest-frame optical
through mid-IR SEDs for a large sample of these so-called ``Hot,
Dust-Obscured Galaxies'' (Hot DOGs). We find that the SEDs of Hot DOGs
are generally well modeled by the combination of a luminous, yet
obscured AGN that dominates the rest-frame emission at $\lambda >
1\mu\rm m$ and the bolometric luminosity output, and a less luminous
host galaxy that is responsible for the bulk of the rest optical/UV
emission. Even though the stellar mass of the host galaxies may be as
large as $10^{11}-10^{12}~M_{\odot}$, the AGN emission, with a range
of luminosities comparable to those of the most luminous QSOs known,
require that either Hot DOGs have black hole masses significantly in
excess of the local relations, or that they radiate significantly
above the Eddington limit, at a level at least 10 times more
efficiently than $z\sim 2$ QSOs. We show that, while rare, the number
density of Hot DOGs is comparable to that of equally luminous but
unobscured (i.e., Type 1) QSOs. This may be at odds with the trend
suggested at lower luminosities for the fraction of obscured AGN to
decrease with increasing luminosity. That trend may, instead, reverse
at higher luminosities. Alternatively, Hot DOGs may not be the
torus-obscured counterparts of the known optically selected, largely
unobscured Hyper-Luminous QSOs, and may represent a new component of
the galaxy evolution paradigm. Finally, we discuss the environments of
Hot DOGs and statistically show that these objects are in regions as
dense as those of known high-redshift proto-clusters.
\end{abstract}


\section{Introduction}

Massive galaxies are thought to evolve from star-forming disks into
passive ellipticals through major mergers that trigger star-formation
and intense episodes of accretion into their central super-massive
black holes (SMBHs) \citep[e.g.][]{hopkins08}. Such a picture can
explain several properties of galaxies, such as the tight correlations
between the mass of their SMBH ($M_{\rm BH}$) and the mass, luminosity
and velocity dispersion of the galaxy's spheroidal component
\citep[e.g.][]{magorrian98,ferrarese00,tremaine02,marconi03,bentz09,gultekin09},
and the evolution of the galaxy luminosity density
\citep[e.g.,][]{faber07}. In these scenarios, the host galaxy stellar
mass is assembled through star-formation ahead of the onset of the
active galactic nuclei (AGN), which through a feedback mechanism heats
up the gas and expels some of it, thereby quenching its
star-formation.

An important characteristic of both the intense star-formation and AGN
episodes is the significant quantities of dust associated with
them. In galaxies undergoing extreme star-formation and AGN activity,
a large fraction of the luminous energy is absorbed by dust and then
re-radiated at infrared/sub-mm wavelengths, as observed for
populations such as Ultra-Luminous Infrared Galaxies
\citep[ULIRGs;][]{sanders96}, Sub-mm Galaxies
\citep[SMGs;][]{blain02,casey14} and Dust-Obscured Galaxies
\citep[DOGs;][]{dey08}. It follows then that studying the most
luminous infrared galaxies in the Universe, which host the most
intense star-formation and AGN activity, likely probe extreme
scenarios within the galaxy evolution paradigm.

NASA's Wide-field Infrared Survey Explorer \citep[WISE;][]{wright10}
was launched in December 2009, and surveyed the entire sky in four
mid-IR bands centered at $3.4\mu\rm m$ (W1), $4.6\mu\rm m$ (W2),
$12\mu\rm m$ (W3) and $22\mu\rm m$ (W4). One of the main goals of the
WISE mission was to identify the most luminous infrared galaxies in
the universe, and the mission has achieved considerable success
towards this goal in the past few years
\citep[][]{eisenhardt12,wu12,bridge13,jones14,stern14,wu14,eisenhardt13,lonsdale14,tsai14}. One
highly successful method to identify such sources is to target objects
that are extremely red in the WISE bands, with faint or no detections
in the more sensitive W1 and W2 bands, but well detected in W3 and
W4. These selection criteria were presented by \citet{eisenhardt12}
and have been shown to successfully identify a population of luminous
galaxies with $z\gtrsim 1$ \citep[for details on the selection
  criteria, see][and \S\ref{ssec:wise_obs}]{eisenhardt12}. Using
sub-mm observations, \citet{wu12} showed that these objects are indeed
extremely luminous, with bolometric luminosities exceeding
$10^{13}~L_{\odot}$, and sometimes exceeding
$10^{14}~L_{\odot}$. These observations also showed that the dust in
these objects is at temperatures significantly higher than in other
luminous infrared populations, such as ULIRGs and SMGs, peaking at
rest-frame wavelengths $\lambda\lesssim 20\mu\rm m$. Such high dust
temperatures are consistent with AGN heating, suggesting the bulk of
the luminosity is produced by accretion onto the galaxy's central SMBH
rather than by star-formation. Indeed, \citet{eisenhardt12} in a
detailed study of one of these sources, WISE J181417.29+341224.9
(W1814+3412 hereafter), determined that its spectral energy
distribution (SED) is consistent with a heavily obscured ($A_V\sim
50~\rm mag$) AGN producing the bulk of the luminosity. Due to their
high dust temperatures, and yet similar optical-to-mid-IR colors to
DOGs, we adopt the terminology of \cite{wu12} and refer to these
objects as Hot, Dust-Obscured Galaxies or Hot DOGs.

Follow-up studies have provided additional interesting aspects of this
population. The observed-frame optical spectra of these objects
\citep{eisenhardt12,wu12,stern14,eisenhardt13} have diverse
properties.  While many of these objects show narrow emission lines
commonly associated with obscured AGN activity present in the IR, some
show features more closely associated with star-formation, with mostly
absorption lines and only Ly$\alpha$ emission \citep[e.g.,
  W1814+3412][]{eisenhardt12}, and some even show red continua with a
lack of emission lines \citep[e.g.,WISE
  J092625.44+423251.9;][]{wu12}. Imaging obtained with the {\it{Hubble
    Space Telescope}}/WFC3 and through adaptive optics with Keck/NIRC2
have shown that these objects typically are not gravitationally
lensed, implying their luminosities are intrinsic
\citep{eisenhardt12,wu14,tsai14}. Recently, \citet{jones14} reported
observations at 850$\mu$m with SCUBA-2 at the James Clerk Maxwell
Telescope (JCMT) of a subsample of 10 Hot DOGs, which suggest these
objects are located in arcmin-scale overdensities of luminous dusty
galaxies (see \S\ref{sec:env_mass}) and confirm the hot dust
temperatures determined by \citet{wu12}. \citet{jones14} constrain the
contribution of ULIRG-type star-formation to less than 30\% of the IR
luminosity, and of spiral-type star-formation to less than
3\%. Similar conclusions are reached by \citet{wu14}, who studied two
Hot DOGs at submm and mm wavelengths at higher spatial resolution
using the Submillimeter Array (SMA) and the Combined Array for
Research in Millimeter-wave Astronomy (CARMA). \citet{wu14} was able
to constrain their cold dust masses to amounts comparable to those of
quasars with comparable luminosities. On the other end of the
electromagnetic spectrum, \citet{stern14} studied the AGN nature of
three Hot DOGs using X-ray observations obtained with the {\it{X-ray
    Multi-Mirror Mission}} ({\it{XMM-Newton}}) and the {\it{Nuclear
    Spectroscopic Telescope Array}} ({\it{NuSTAR}}), finding that the
AGN emission is heavily absorbed, possibly Compton-thick. Using a
similar sample to that defined by the Hot DOG selection criteria,
\citet{bridge13} determined that a significant fraction of such
objects show extended Ly$\alpha$ emission on 30 -- 100~kpc scales, and
pointed out this could be consistent with the presence of intense
quasar feedback. \citet{bridge13} also further constrained the high
temperature of these objects by using {\it{Herschel}}/PACS and SPIRE
observations to map the full shape of their far-IR SEDs. Finally,
\citet{lonsdale14} present a study based on ALMA Cycle 0 observations
of the far-IR SEDs of radio-selected, red WISE objects that are
possibly the radio-loud counterparts of Hot DOGs. Although these
objects are located at somewhat lower redshifts (0.47--2.85), they
share some of the same characteristics, including the overall high
dust temperatures and an even more dramatic apparent overdensity of
nearby luminous dusty galaxies \citep{jones14b}.

In this work we study the physical properties of Hot DOGs by analyzing
their SEDs, number densities and environments. In a companion paper,
\citet{tsai14} presents a detailed study of the most luminous Hot
DOGs, those with bolometric luminosities in excess of
$10^{14}~L_{\odot}$. The article is structured as follows. In Section
\ref{sec:data} we discuss the sample selection and the follow-up
photometric and spectroscopic observations. In Section
\ref{sec:sed_fits} we present our SED modeling methodology, while in
Section \ref{sec:analysis} we apply it to model our sample of Hot DOGs
and discuss their inferred physical properties. In Section
\ref{sec:qso_hd_comp} we compare the number density of Hot DOGs to
that of comparably luminous QSOs. Finally, in Section
\ref{sec:env_mass} we study the density of the environments in which
Hot DOGs are found using follow-up {\it{Warm Spitzer}}/IRAC
imaging. We discuss how the environments compare to known clusters at
similar redshifts, and how this constrains the stellar masses of Hot
DOGs. Throughout this work we assume a flat $\Lambda$CDM cosmology
with $H_0 = 73~\rm km~\rm s^{-1}$, $\Omega_M = 0.3$, and
$\Omega_{\Lambda} = 0.7$. We refer to all magnitudes in the Vega
photometric system. For convenience, the different samples used
throughout this work are summarized in Table \ref{tab:samples}.

\input{tab1}

\section{Sample Selection and Multi-Wavelength Follow-Up Observations}\label{sec:data}

\subsection{WISE and the W12drop Selection}\label{ssec:wise_obs}

The WISE mission observed the full sky in four mid-IR photometric
bands with a FWHM of 6\arcsec\ in W1--3 and 12\arcsec\ in W4. We use
the WISE All-Sky data release, which includes all observations
obtained during the fully cryogenic mission. WISE surveyed the sky in
a polar orbit with respect to the ecliptic, simultaneously obtaining
images in all four bands. Hence, the number of observations in a field
increases with its ecliptic latitude. While fields near the ecliptic
were typically observed 12 times, the number can grow to thousands
near the ecliptic poles \citep[e.g.,][]{jarrett11}. The median
coverage across the sky is approximately 15 frames per
passband. Detailed accounts of the mission are presented by
\citet{wright10} and in the WISE All-Sky data release explanatory
supplement\footnote{\url{http://wise2.ipac.caltech.edu/docs/release/allsky/expsup/}}.

As discussed earlier, our canonical picture of galaxy evolution
suggests the existence of key stages where massive galaxies experience
extremely luminous but heavily dust-enshrouded star-formation and
nuclear activity. For the most massive galaxies, these stages may
reach infrared luminosities $L_{\rm IR}>10^{13}~L_{\odot}$, and hence
be classified as Hyper-Luminous Infrared Galaxies (HyLIRGs), but can
be very faint in the optical bands due to obscuration. \citet{wu12}
and \citet{eisenhardt12} presented a large sample of WISE-selected
HyLIRGs, which are the main target of this study. The selection
criteria used by \citet{eisenhardt12} and \citet{wu12} specifically
target galaxies red enough to be well detected in the long wavelength
WISE bands W3 and W4, but are poorly or undetected at the shortest
wavelength, more sensitive W1 and W2 bands. Samples selected in this
way are referred to as ``W1W2-dropouts'' by \citet{eisenhardt12} and
\citet{wu12}, but for brevity we use ``W12drops'' here as an
equivalent term.

W12drop selection requires that $\rm W1>17.4~\rm mag$, and that
either
\begin{equation}\label{eqn:w12d_sel}
  \rm W4\ <\ 7.7~\rm mag\ \wedge\ \rm W2 - \rm W4\ >\ 8.2~\rm mag,
\end{equation}
\noindent or
\begin{equation}
  \rm W3\ <\ 10.6~\rm mag\ \wedge\ \rm W2 - \rm W3\ >\ 5.3~\rm mag.
\end{equation}
\noindent Furthermore, objects are required to be farther than
$30^{\circ}$ from the Galactic Center and $10^{\circ}$ from the
Galactic Plane to limit contamination by Galactic objects. All objects
are required to be free of artifacts flagged by the WISE pipeline and
to not be associated with either known asteroids or those discovered
by WISE \citep[][]{mainzer11}.

Finally, we required candidates to pass a series of visual inspections
of both individual exposures and coadded images for any given
source. We focused such efforts on the brighter candidates with $\rm
W4<7.2$, resulting in a sample of 252 objects over approximately
$32,000~\rm deg^2$. We refer to these objects as the ``core
sample''. A search to W4$\lesssim$ 7.7 using preliminary reductions
covering 70\% of the total area was also carried out, resulting in an
additional sample of 682 W12drops which is somewhat less complete and
well-characterized than the core sample. We refer to the total sample
of 934 objects as the ``full sample''.

It is now known that WISE All-Sky profile-fitting derived fluxes of
very faint sources are significantly biased due to excess sky
subtraction during the data processing \citep[see][]{lake13}. The
effect is somewhat stochastic in nature, but can be well modeled as a
constant underestimation of $9.29\pm 0.04~\mu\rm Jy$ and $10.38\pm
0.07~\mu\rm Jy$ in the W1 and W2 fluxes of the WISE All-Sky release
catalog (S.~Lake, private communication). Some of these issues have
been corrected in the latest WISE data release, dubbed AllWISE, but
because the All-Sky Catalog was used for the W12drop selection we use
the WISE All-Sky fluxes and apply the corrections outlined above when
modeling the selection function rather than trying to translate the
W12drop selection function to the AllWISE data release. We note that
the SED modeling discussed later is not affected by this issue, since
we rely on deeper {\it{Warm Spitzer}} observations for those
wavelengths (see \S\ref{ssec:spitzer_obs} and \S\ref{sec:analysis} for
details), but it will prove to be important when considering the
W12drop selection function (see \S\ref{sec:qso_hd_comp} for
details).

Additionally, \citet{wright10} found systematic differences in the W3
and W4 magnitudes of red and blue calibrators, with the red
calibrators being 17\% too faint in W3 and 9\% too bright in W4. Given
the red colors of our sources, we have corrected their W3 and W4
magnitudes by adding --0.17~mag and 0.09~mag respectively to the
quantities reported in the All-Sky release for modeling their SEDs
(see \S\ref{sec:sed_fits}). Since these corrections are not considered
by the selection criteria, we remove them when evaluating the W12drop
selection function in \S\ref{sec:qso_hd_comp}. \citet{brown14} have
recently suggested a somewhat larger correction of 0.13~mag instead
0.09~mag for sources as red as those considered here. Our main results
would not be qualitatively affected by using this larger correction.

\subsection{Follow-up Observations}

\subsubsection{{\it{Warm Spitzer}} Observations}\label{ssec:spitzer_obs}

We obtained observations of Hot DOGs with the IRAC instrument
\citep{fazio04} onboard the {\it{Spitzer Space Telescope}}
\citep{werner04}. In its non-cryogenically cooled state, known as
      {\it{Warm Spitzer}}, the IRAC camera obtains photometry in two
      broad-band channels centered at $3.6$ and $4.5\mu$m, referred to
      as [3.6] and [4.5]. The channels are similar to the WISE W1 and
      W2 bands, but because of its larger aperture, longer exposure
      time, and smaller PSF (FWHM of $\approx$1.7\arcsec\ in each
      band), IRAC provides significantly deeper images. We refer the
      reader to \citet{griffith12} for details of this program, as
      well as of the data reduction and photometric measurements.

Of the 934 W12drops in the full sample, 712 were observed with
{\it{Warm Spitzer}} in the [3.6] and [4.5] IRAC channels. All but one
of these is well detected in both bands, with the one outlier object
(W0149--8257) detected only in the [4.5] band. We limit our parent
sample to those 711 objects detected in both {\it{Warm Spitzer}}/IRAC
bands. The core sample (W4$<$7.2) is similarly reduced to 103 targets.

\subsubsection{Ground-Based Near-IR Imaging}

We obtained follow-up near-IR observations of our sample using the
Wide-field IR Camera \citep[WIRC;][]{wilson03} on the Hale 200-inch
telescope at Palomar Mountain, the WIYN High-Resolution Infrared
Camera \citep[WHIRC;][]{meixner10} at the 3.5m WIYN telescope, the
Ohio State InfraRed Imager/Spectrometer \citep[OSIRIS;][]{depoy93} at
the 4m SOAR telescope, and the SAO Widefield InfraRed Camera
\citep[SWIRC;][]{brown08} at the 6.5m MMT telescope. Table
\ref{tab:nir_mags} provides more details about these observations.

\input{tab2}

All images were reduced following standard IRAF procedures using the
XDIMSUM
package\footnote{\url{ftp://iraf.noao.edu/ftp/extern-v214/xdimsum}},
and all fluxes were obtained in 4.0\arcsec\ diameter apertures. Each
image was flux-calibrated using the 2MASS point source catalog
\citep[PSC,][]{skrutskie06}, using comparison stars within the field
of view whenever possible, or by using the closest observation in time
of a field containing 2MASS detected stars if conditions were
photometric. The latter was only necessary for some of the OSIRIS
observations, which have an 80\arcsec\ field of view. We add the
dispersion of the zero-point calibration in quadrature to the
photometric uncertainty of each source. Magnitudes are listed in Table
\ref{tab:nir_mags} for each of the W12drops for which we obtained
follow-up near-IR observations.

Of the 711 (103) objects in our full (core) W12drop sample with
{\it{Warm Spitzer}} observations, 84 (52) have been observed in
$J$-band, 23 (16) in $H$-band, and 37 (19) in $K$-band. Of these, only
1 (0) object has been observed in all three bands, and only 26 (16)
have been observed in more than one band.

\subsubsection{Optical Spectroscopy}\label{sssec:opt_spec}

\begin{figure}
  \begin{center}
    \plotone{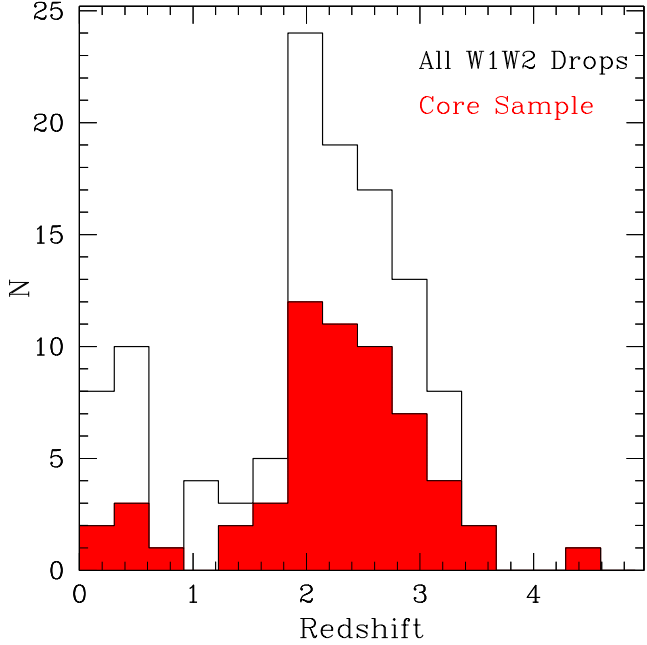}
    \caption{Redshift distribution of the 115 W12drops with
      spectroscopic redshift and IRAC measurements. The shaded
      histogram shows the distribution for the 58 objects in the core
      sample (W4$<$7.2, see \S\ref{ssec:wise_obs}).}
    \label{fg:zhist}
  \end{center}
\end{figure}

Optical spectroscopy was performed for a large fraction of our sample
using several facilities, with greater emphasis given to the core
sample. These were obtained primarily using LRIS on the Keck-I
telescope \cite[see][for examples]{wu12} and GMOS-S on the Gemini-S
telescope. Redshifts are generally based on multiple features and are
therefore considered secure. \citet{eisenhardt13} will provide a
comprehensive description of the optical spectroscopy. Figure
\ref{fg:zhist} shows the redshift distribution of all 115 (58) objects
in the full (core) W12drop sample with {\it{Warm Spitzer}}
observations and successful redshift measurements. The full sample
shows a clear minimum at $0.6 < z < 2$, suggesting the low- and
high-redshift populations are distinct. The core sample distribution
is also consistent with the presence of the bimodality, albeit at
lower significance due to the smaller number of objects. Since we are
only interested in very luminous objects, we focus on the sample of 96
(52) objects from the full (core) sample with
$z>1$. \citet{eisenhardt13} shows that approximately 70\% of the
objects targeted for spectroscopy yielded redshift measurements, but
that objects with failed spectroscopic measurements are primarily due
to optical faintness, suggesting they are typically located at high
redshift and are bona-fide HyLIRGs. We note that during some of the
spectroscopic observing runs, we biased against targets detected in
both the $B$ and $R$ bands of the Digitized Sky
Survey\footnote{\url{https://archive.stsci.edu/cgi-bin/dss\_form}} to
reduce the incidence of low redshift contaminants. This bias has only
a minor effect for the $z>1$ population, so we do not discuss it any
further.

\section{SED Modeling Methodology}\label{sec:sed_fits}

In order to gain insight into the physical properties of Hot DOGs, we
study their rest-frame optical through mid-IR multi-wavelength
SEDs. Specifically, we study their rest-frame optical through mid-IR
properties by combining the WISE data with optical spectroscopy,
{\it{Warm Spitzer}} observations and ground-based near-IR
photometry. We model the SEDs following the approach applied in
\citet{eisenhardt12} to study Hot DOG W1814+3412. Namely, we use the
low resolution AGN and galaxy SED templates and the respective
modeling algorithm of \citet{assef10}. Briefly, every object is
modeled as a non-negative linear combination of three host galaxy SED
templates \citep[broadly resembling E, Sbc and Im types, see][for
  details]{assef10} and one AGN SED template. \citet{eisenhardt12}
used the AGN template of \citet{richards06} because of its longer
wavelength range. Here we use the \citet{assef10} AGN template
instead, but there is no substantial difference between these AGN
templates for the purpose of the SED modeling of Hot DOGs over the
UV-through-mid-IR wavelength range studied in this work.

We also fit an AGN reddening component, which we parametrize by the
color excess $E(B-V)$, considering values from 0 to $10^{1.5}$. The
assumed reddening-law corresponds to an SMC-like extinction for
$\lambda<3300$\AA, and a Galactic extinction curve at longer
wavelengths. Additional details are provided in
\citet{assef10}. Although in some unresolved obscured sources it may
be important to modify these reddening curves to account for the
optical and UV photons scattered into the line-of-sight
\citep[see][]{kochanek12}, our sources are under too much obscuration
for this to be a significant issue. In the next sections we show that
this approach does, in general, perform a good job of modeling Hot
DOGs, although possible shortcomings are discussed in detail.

One of the important quantities we want to study is the intrinsic
bolometric luminosity of the underlying AGN. We estimate it using the
scaling relation of \citet{kaspi00},
\begin{equation}\label{eq:lbol}
  L_{\rm AGN}^{\rm Bol}\ =\ 9\ \lambda L_{\lambda}(5100\rm \AA).
\end{equation}
\noindent The continuum luminosity at 5100\AA\ is calculated by taking
only the best-fit reddened AGN component, and removing the
obscuration. While a more self-consistent bolometric luminosity can be
obtained by integrating over the best-fit AGN component to the SED
\citep[see, e.g.,][]{assef10,assef13,eisenhardt12}, this scaling is
widely used, so adopting it simplifies the comparison with other
results in the literature. For reference, the luminosity obtained
integrating over the unreddened AGN template between 0.1 and 30$\mu$m
is greater than the bolometric luminosity estimated using equation
(\ref{eq:lbol}) by a factor of 1.3. By using a single AGN template we
are implicitly assuming the torus covering fraction of the accretion
disk in all Hot DOGs is equal to that implied by the template, which
is likely about 50\% given the results of \citet{assef13}. For
luminous Type 1 quasars \citet{roseboom13} found that the mean
covering fraction was 39\% with a dispersion of 18\%. Since the
amplitude of the AGN component in the best-fit SED is primarily
anchored by the W3 and W4 fluxes, a larger covering fraction could
reduce the deduced AGN continuum luminosity at 5100\AA. In principle,
this luminosity could be overestimated by up to a factor of 2 in the
extreme scenario where the torus leaves no open line of sight towards
the accretion disk and is composed of uniformly distributed hot dust
clouds, as all dust must strongly radiate at $\lambda \lesssim
10\mu\rm m$ to affect our results. This scenario is, however,
unlikely, given the detection of AGN narrow-emission lines in many of
the rest-frame UV spectra \citep[see, e.g.][]{wu12}.

We also wish to estimate the stellar mass ($M_{*}$) of each host
galaxy. We estimate this quantity by multiplying the rest-frame
luminosity of the host component in the $K$-band by the mass-to-light
ratio ($M/L$) in that band. The value of $M/L$ depends on many
parameters, including the galaxy's star-formation history,
metallicity, stellar initial mass function (IMF) and contribution from
thermally pulsating asymptotic giant branch (TP-AGB) stars. Because we
generally have only a single photometric band probing the host
properties, we only aim to place meaningful bounds on $M/L$. Although
the lower end of the $M/L$ range is only loosely bound, the upper end
is much better constrained, as it is primarily limited by the age of
the Universe at the redshift of the object. Hence, for the purpose of
this study, we will focus on estimating upper bounds on the stellar
mass of each Hot DOG. We estimate these upper bounds for each object
using the EzGal code of \citet{mancone12a} with the stellar population
models of \citet{bruzual03}. We choose these stellar population models
in favor of more recent ones available for EzGal
\citep[e.g.,][]{maraston05,conroy09,conroy10} since they have the
lowest contribution of TP-AGB stars to the composite SEDs, implying
the highest $M/L$ values. The $M/L$ values in the rest-frame near-IR
are rather insensitive to variations in the metallicity, with lower
metallicities implying higher $M/L$ values in $K_s$. To be
conservative we consider the lowest metallicity available for EzGal,
$Z=0.008~(\equiv 0.4 Z_{\odot})$. For the star-formation history, we
consider a simple stellar population (SSP) with a formation redshift
$z_F = 15$. Finally, for the IMF, we consider the results of
\citet{conroy13}, who have shown that in early-type galaxies, the
$M/L$ ratio in $K$-band can be up to twice that expected for the Milky
Way. We use $M/L$ values two times higher than those estimated
assuming a \citet{chabrier03} IMF. Our only assumption that would tend
to underestimate the stellar mass is that of little host obscuration
in the rest-frame $K$-band. We discuss this issue further in detail in
\S\ref{sec:analysis} and \S\ref{sec:env_mass}, where evidence is shown
that although Hot DOGs live in significantly dense regions, possibly
clusters or dense filaments, higher stellar masses would predict even
richer environments than observed.

Finally, we also attempt to estimate the central SMBH mass ($M_{\rm
  BH}$) for these objects. In many cases, however, we do not quote the
values of $M_{\rm BH}$ but of the Eddington luminosity defined as
\begin{equation}\label{eq:eddington_luminosity}
  L_{\rm Edd}\ =\ 3.28\times 10^4\ \left(\frac{M_{\rm
      BH}}{M_{\odot}}\right)\ L_{\odot},
\end{equation}
\noindent which corresponds to the luminosity at which photon pressure
inhibits isotropic accretion onto an isotropically radiating
body. This is an interesting quantity to study for AGN since most
energy is generated by accretion onto the SMBH. Furthermore,
\citet[][also see \citealt{shen08}]{kollmeier06} have shown that
luminous QSOs at similar redshifts as Hot DOGs tend to radiate in a
limited range of Eddington ratios, defined as $\lambda_E = L_{\rm
  AGN}^{\rm Bol}/L_{\rm Edd}$.

Direct estimates of $M_{\rm BH}$ in AGN, and hence $L_{\rm Edd}$,
based on single optical spectra are possible by combining the width of
their broad emission lines and the luminosity of their accretion disks
\citep[e.g.,][]{vestergaard06}, typically limiting such measurements
to unobscured objects. In an upcoming article \citep{wu14b} we explore
such estimates for a handful of Hot DOGs where we observed broad
H$\alpha$ in the near-IR, but such methods are certainly not
applicable to the sample of objects we study here. Hence, we consider
two alternative methods, and explore their consequences in depth in
\S\ref{sec:analysis}. First, we estimate $M_{\rm BH}$ through equation
(\ref{eq:eddington_luminosity}) by assuming that the AGN in Hot DOGs
radiate at the typical $\lambda_E=$ 0.30 determined by
\citet{kollmeier06} for QSOs at similar redshifts. Alternatively, we
assume that $M_{\rm BH}$ is related to the stellar mass in the host
galaxy in the same way as found for local galaxies. Specifically, we
use the relation between the spheroidal component mass ($M_{\rm Sph}$)
and $M_{\rm BH}$ of \citet{bennert11a},
\begin{equation}
\log \frac{M_{\rm BH}}{M_{\odot}}\ =\ -3.34\ +\ 1.09 \log \frac{M_{\rm
    Sph}}{M_{\odot}}.
\end{equation}
Given that the near-IR imaging reported here is either ground based or
from {\it{Spitzer}}, we cannot attempt a morphological decomposition
of the bulge or spheroidal component. Furthermore, due to the extreme
nature of Hot DOGs, it is not even clear the definition of a
spheroidal component would be sensible. Hence, we assume that all of
the detected rest-frame $K$-band light belongs to the spheroidal
component. Combined with the fact that our stellar mass estimates are
also upper bounds, our estimates of $M_{\rm BH}$ obtained in this
manner should be considered as generous upper limits. In the next
section we discuss how both estimates lead to very different scenarios
for Hot DOGs, highlighting that accurate estimates of $M_{\rm BH}$ in
Hot DOGs are a crucial element for understanding their nature,
requiring more extensive optical/near-IR spectral coverage.

Due to the low number of degrees of freedom, there can be considerable
uncertainty for the best-fit parameters of each galaxy. To account for
this, we do a Monte Carlo approach. For each object we resample each
available flux from a Gaussian distribution centered at the measured
value with a standard deviation equal to the photometric uncertainty,
and re-estimate all the parameters described above. We repeat this
1000 times per galaxy and re-estimate the results discussed in the
next section, finding no significant difference.

Finally, we caution that the infrared emission might not be powered by
AGN activity but might be powered by extreme starburst disks, such as
proposed by \citet{thompson05}, who found that, under the appropriate
conditions, a disk of star-formation can form where gas and dust are
supported primarily by star-formation feedback and radiation
pressure. The typical dust temperature can become quite high,
comparable to the temperatures of $\sim 60~\rm K$ typical of Hot DOGs
\citep{wu12}. Extreme star-formation rates, on the order of $\sim
10^4~\rm M_{\odot}~\rm yr^{-1}$, would be needed to power the
luminosities seen in Hot DOGs \citep[see, e.g.][]{eisenhardt12},
making this scenario quite unlikely. Also, \citet{thompson05} find
that it is inevitable to form a bright AGN at the center of such a
starburst disk and that, furthermore, the star-formation driven
rest-frame 10$\mu$m emission still relates to the accretion disk
emission in the same way as for a regular Type 1 QSO as judging from
the templates of \citet{elvis94}. Hence, even in this arguably
unlikely physical scenario, the main results presented here are still
valid. We do not refer to this scenario hereafter.

\section{Analysis}\label{sec:analysis}

\begin{figure}
  \begin{center}
    \plotone{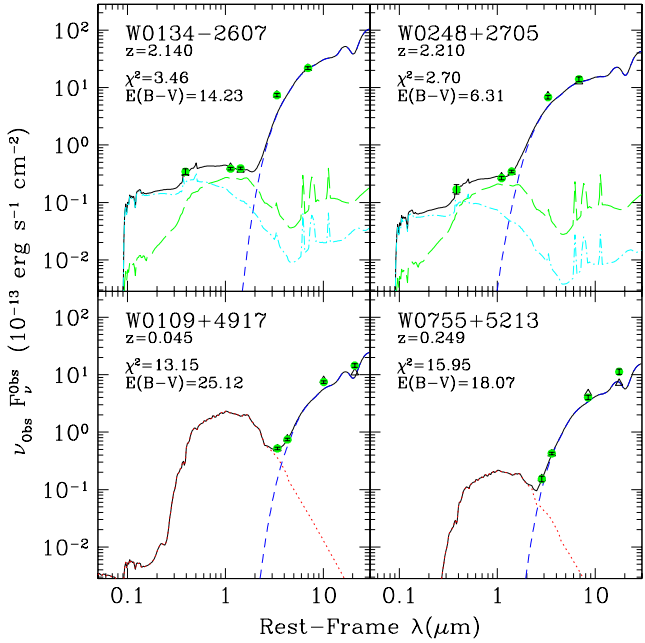}
    \caption{Examples of SED fits to two high-redshift (top panels)
      and two low-redshift W12drops (bottom panels). All four have
      $\chi^2$ values close to the median values of their
      populations. Each panel shows the observed fluxes ({\it{green
          filled circles}}), as well as the expected fluxes from the
      best-fit SED model ({\it{black open triangles}}). The solid lack
      line shows the best-fit SED, composed of an AGN under $E(B-V)$
      of obscuration ({\it{dashed blue line}}) plus a quiescent
      ({\it{dotted red}}), intermediate ({\it{long-dashed green
          line}}) and strongly star-forming ({\it{dot-dashed cyan
          line}}) stellar populations (see \S\ref{sec:sed_fits} for
      details). Most galaxies do not require all three stellar
      templates. Notice the much higher $\chi^2$ values of the
      low-redshift galaxy fits.}
    \label{fg:sed_examples}
  \end{center}
\end{figure}

By design, W12drops are either undetected or marginally detected in W1
and W2. Because of this, we fit their SEDs using the {\it{Warm
    Spitzer}} [3.6] and [4.5] photometry, W3 and W4 photometry, and,
whenever possible, the ground-based near-IR photometry. Figure
\ref{fg:sed_examples} shows examples of SED fits to two high-redshift
and two low-redshift W12drops. Objects with $z<1$ have much lower
luminosities and are much worse fit by the models than their
higher-redshift counterparts, reinforcing the idea they compose two
distinct populations. From now on, we will use interchangeably the
terms ``Hot DOG'' \citep{wu12} and ``W12drop at $z>1$'' for
convenience in this article. 

\begin{figure}
  \begin{center}
    \plotone{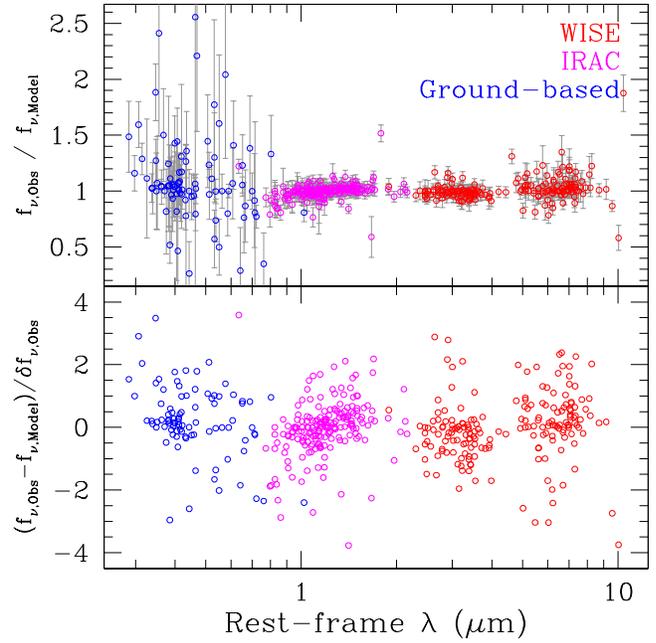}
    \caption{Flux residuals of the SED models for the 89 out 96 Hot
      DOGs with $\chi^2<20$.}
    \label{fg:residuals}
  \end{center}
\end{figure}

Our SED fits show that the AGN component dominates the luminosity of
Hot DOGs, accounting for $>$97\% of the total $0.1-30\mu\rm m$ output
in all objects. Figure \ref{fg:residuals} shows the residuals of the
SED fits both as a ratio between the observed and model fluxes (top
panel) and as the difference with respect to the measurement
uncertainties (bottom panel). The ground-based near-IR photometry
shows large discrepancies with respect to the modeled fluxes but only
in absolute terms, as the deviations are in almost no case beyond
3$\sigma$. The {\it{Warm Spitzer}}/IRAC and W3-4 fluxes show only
small absolute discrepancies from the model, without a significant
systematic component. The lack of systematic residuals also suggests
it is unlikely there is a significant amount of unrecognized
obscuration on the host galaxy that would bias our maximal stellar
mass estimates (see \S\ref{sec:sed_fits}).

\begin{figure}
  \begin{center}
    \plotone{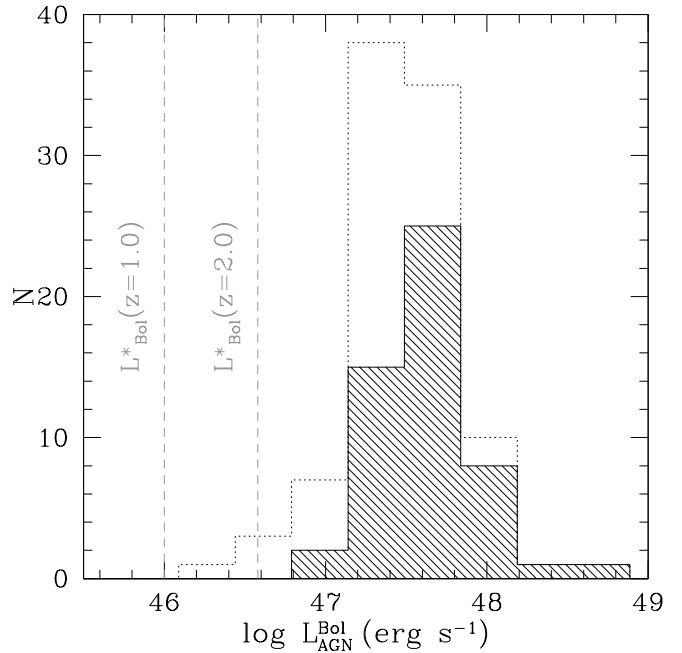}
    \caption{Distribution of AGN bolometric luminosities derived from
      our SED modeling of Hot DOGs (see \S\ref{sec:sed_fits} and
      \S\ref{sec:analysis} for details). The solid hashed histogram
      shows the distribution of the core sample (i.e., W4$<$7.2),
      while the dashed open one shows the distribution of the full
      sample. The vertical dashed gray lines show the characteristic
      AGN bolometric luminosity derived from the $J$-band QLF of
      \citet{assef11}, assuming the unobscured AGN SED template of
      \citet{assef10} and equation (\ref{eq:lbol}).}
    \label{fg:agn_lbol}
  \end{center}
\end{figure}

Figure \ref{fg:agn_lbol} shows the distribution of the bolometric
luminosity of the AGN component $L_{\rm AGN}^{\rm Bol}$ for Hot DOGs
in the full sample as well as in the core sample. Hot DOGs tend to
have quite luminous AGN, with bolometric luminosities between
$10^{47}$ and $10^{48}~\rm erg~\rm s^{-1}$. In the same panel, we show
the values of the characteristic quasar luminosity function (QLF)
luminosity $L^*$ at two different redshifts, $z=1.0$ and 2.0,
determined by converting appropriately the values of $M_{*,J}$
determined by \citet{assef11}. The AGN components in Hot DOGs are
among the most luminous AGN at their redshifts. We discuss this
further in \S\ref{sec:qso_hd_comp}.

\begin{figure}
  \begin{center}
    \plotone{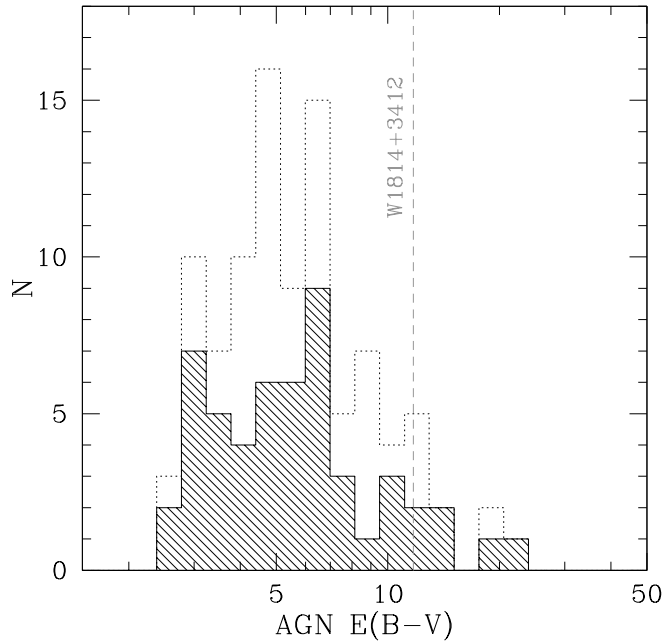}
    \caption{Distribution of AGN obscurations derived from our SED
      modeling of Hot DOGs (see \S\ref{sec:sed_fits} and
      \S\ref{sec:analysis} for details). The solid hashed histogram
      shows the distribution of the core sample (i.e., W4$<$7.2),
      while the dashed open one shows the distribution of the whole
      sample. The dashed gray line shows the obscuration derived for
      HyLIRG W1814+3412, studied by \citet{eisenhardt12}. For
      reference, we note that for the reddening law used in this
      article, the attenuation factor at 1$\mu$m is given by $A_{1\mu
        m} = 1.24*E(B-V)$.}
    \label{fg:ebv}
  \end{center}
\end{figure}

Figure \ref{fg:ebv} shows the distribution of dust obscuration towards
the accretion disk, parametrized by $E(B-V)$. \citet{eisenhardt12}
studied Hot DOG W1814+3412 in detail using 8 bands of optical through
mid-IR photometry, finding an obscuration of $E(B-V) = 15.6\pm1.4$
with a very similar approach to that used here. With an updated
processing of the WISE data and the corrections to W3 and W4 discussed
in \S\ref{ssec:wise_obs} but a more limited multi-wavelength
photometry set than that used by \citet{eisenhardt12}, we find here a
lower obscuration of $E(B-V)=11.7\pm 1.2$ for W1814+3412. This
obscuration is lower primarily due to the W3 and W4 band corrections
applied here. If we do not consider these corrections, we find an
obscuration of $E(B-V)=15.1\pm 1.2$ for W1814+3412, consistent with
the value found by \citet{eisenhardt12}. Figure \ref{fg:ebv} shows
that W1814+3412 has significantly more obscuration than the average
Hot DOG, for which $\langle E(B-V)\rangle = 6.8$ in the core sample
and $\langle E(B-V)\rangle = 6.4$ in the full sample. For
completeness, we note the median obscuration is $E(B-V) = 6.0$ in the
core sample and $E(B-V) = 5.5$ in the full sample. The best-fit AGN
obscuration for Hot DOGs ranges between $2.5 < E(B-V) <
21.5$. Assuming the median gas-to-dust ratio of \citet{maiolino01},
$E(B-V)/N_{\rm H} = 1.5\times 10^{-23}~\rm cm^{2}~\rm mag$, the
average obscuration corresponds to gas column densities of
approximately of $4\times 10^{23}~\rm cm^{-2}$, and a range of
$1.6\times 10^{23} < N_{\rm H} < 1.4\times 10^{24}~\rm cm^{-2}$. This
is over 10 times more absorption than the typical dividing line
between Type 1 and Type 2 AGN \citep[$N_{\rm H} = 10^{22}~\rm
  cm^{-2}$; see, e.g.,][]{ueda03}, to just being slightly below
Compton thick ($N_H > 1.5\times 10^{24}~\rm cm^{-2}$). Using X-ray
observations of a sample of three Hot DOGS obtained with
{\it{XMM-Newton}} and {\it{NuSTAR}}, \citet{stern14} inferred for each
of them neutral hydrogen column densities of $N_{\rm H} \gtrsim
10^{24}~\rm cm^{-2}$, consistent with Compton-thick obscuration, and
in good agreement with the values inferred above from dust
obscuration.

\begin{figure}
  \begin{center}
    \plotone{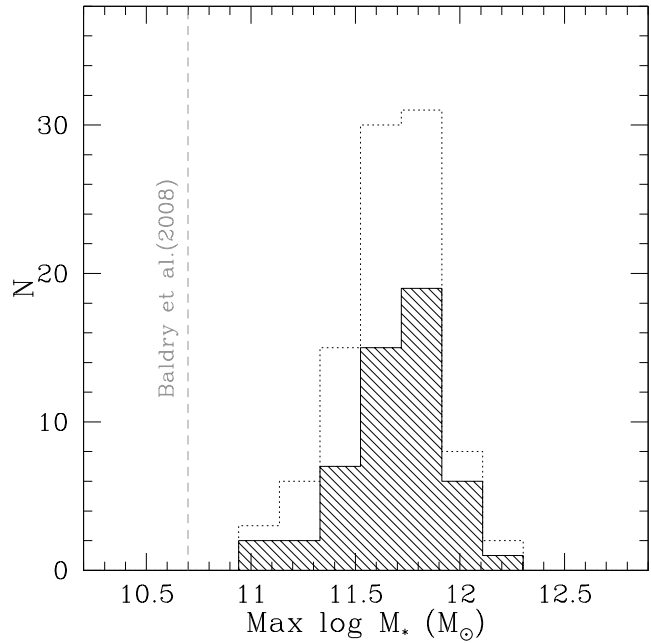}
    \caption{Distribution of stellar mass upper bounds derived from
      our SED modeling of Hot DOGs (see \S\ref{sec:sed_fits} and
      \S\ref{sec:analysis} for details). The solid hashed histogram
      shows the distribution of the core sample (i.e., W4$<$7.2),
      while the dashed open one shows the distribution of the whole
      sample. The dashed gray line shows the stellar mass of an $L*$
      galaxy today from \citet{baldry08}.}
    \label{fg:max_mstar}
  \end{center}
\end{figure}

Figure \ref{fg:max_mstar} shows our maximal stellar mass
estimates. For reference, we also show the stellar mass of an $L*$
galaxy today \citep[$\sim 5\times 10^{10}~\rm M_{\odot}$;
  e.g.,][]{baldry08}. Even if we were to assume a mass-to-light ratio
10 times lower, as appropriate for extreme starbursts, Hot DOGs would
still have massive host galaxies. In \S\ref{sec:env_mass} we further
discuss these host mass estimates and explore their environments in
the context of such massive host galaxies. However, as massive as the
host galaxies are, the AGN still dominates the emission by orders of
magnitude, which has very interesting implications for the nature of
Hot DOGs depending on the mass of their SMBH. As discussed earlier,
however, we are not able to obtain unique estimates of $M_{\rm
  BH}$. Below we explore two scenarios for which we can estimate
$M_{\rm BH}$ based on indirect considerations as well as the
implications of each scenario to the nature of Hot DOGs.

\subsection{$M_{\rm BH}$ Estimates based on a Fixed $\lambda_E$}\label{ssec:mbh_case1}

\begin{figure}
  \begin{center}
    \plotone{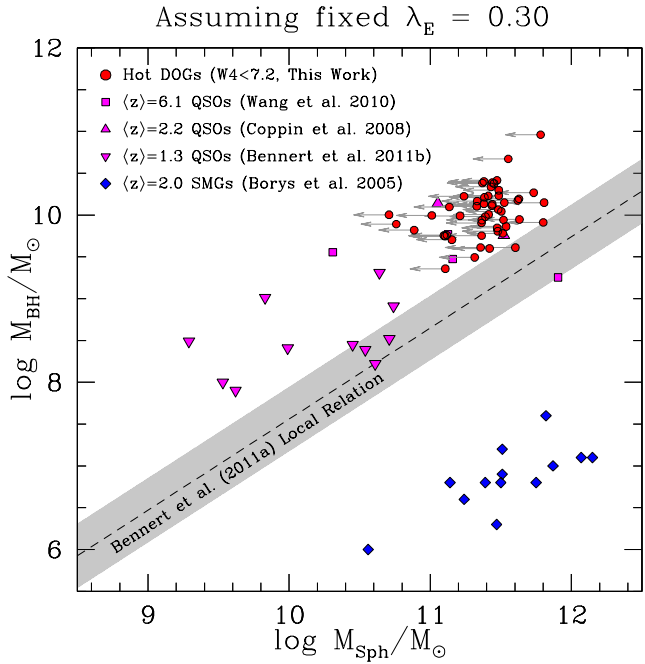}
    \caption{The position in the $M_{\rm BH}-M_{\rm Sph}$ diagram of
      Hot DOGs ({\it{red circles}}) if we assume the AGN in Hot DOGs
      radiate at the typical $\lambda_E=0.30$ of $z\sim 2$ QSOs. For
      comparison, we also show the relation of local active galaxies
      determined by \citet{bennert11a}, as well as the position in the
      diagram of different high redshift populations in the
      literature, as described in the top left labels. For the
      literature samples, the label shows their respective mean
      redshifts.}
    \label{fg:mbh_msph}
  \end{center}
\end{figure}

The first scenario estimates $M_{\rm BH}$ assuming a fixed Eddington
ratio of $\lambda_E =$ 0.30, the same Eddington ratio as that
determined by \citet{kollmeier06} for $z\sim 2$ QSOs. Since the AGN
emission is so luminous, a very large $M_{\rm BH}$ is needed to have
an SMBH accreting at the same level of regular QSOs. In fact, the SMBH
mass is much larger than expected for the local stellar bulge to
$M_{\rm BH}$ ratio \citep{bennert11a} and the stellar mass of the host
galaxy, even for the upper limits calculated above, as shown in Figure
\ref{fg:mbh_msph}. Even if we assume that the stellar mass is equal to
the upper bound and is dominated by an spheroidal component, Hot DOGs
sit an order of magnitude above the local relation, but the
discrepancy is likely much larger. If these galaxies are to evolve in
such a way that at $z=0$ they would fall in the local measured
relations, the bulge will have to grow at a much faster rate than the
accreting central SMBH, implying that SMBH growth precedes the host
galaxy growth. In such a scenario, the stellar mass growth could not
be quenched by intense AGN activity, as is assumed in many evolution
models of massive galaxies \citep[as in, e.g.,][]{hopkins08}, since
the epoch of intense SMBH growth significantly precedes the end of the
host galaxy assembly, but may be consistent with quenching by
low-level AGN accretion through radio-mode feedback \citep[as in,
  e.g.,][]{croton06}.

While this discrepancy can be alleviated by considering a larger value
for $\lambda_E$, it is consistent with what has been found previously
for some Type 1 QSOs. For example, \citet{bennert11b} studied the
evolution of the $M_{\rm BH}-M_{\rm Sph}$ relation with cosmic time
using a sample of $z\sim 1$ unobscured QSOs observed with the
{\it{Hubble Space Telescope}} where $M_{\rm Sph}$ was determined
through SED and morphological modeling of the host. \citet{bennert11b}
found a trend in the same direction suggested by our data, namely
$M_{\rm BH}/M_{\rm Sph}\propto (1+z)^{1.96\pm0.55}$, while combination
with more heterogeneous measurements in the literature yielded $M_{\rm
  BH}/M_{\rm Sph}\propto (1+z)^{1.16\pm0.15}$. Figure
\ref{fg:mbh_msph} shows the set of objects studied in detail by
\citet{bennert11b}, illustrating a displacement from the local
relation comparable to that of Hot DOGs. Similar results have also
been obtained for unobscured QSOs at $z\sim 6$ by \citet[][and
  references therein]{wang10} and at $z\sim 2$ by \citet{coppin08},
where the host stellar masses come from dynamical estimates based on
CO emission and the assumption of random orientation (i.e.,
$i=32.7^o$). These findings support the assumption that the AGN in Hot
DOGs radiate at the same $\lambda_E$ as similar redshift QSOs. For
comparison, Figure \ref{fg:mbh_msph} also shows the location in this
diagram of $z\sim 2$ SMGs, as determined by \citet{borys05}, which are
highly discrepant from the local relation but in the opposite
direction to QSOs. Interestingly, \citet{bongiorno14} has recently
shown that $z\sim 2$ Type 1.8/1.9 QSOs may actually show better
agreement with the local relation than Type 1 QSOs, with a discrepancy
that may depend on $M_{\rm BH}$.

A different picture to interpret these results is that proposed by
\citet[][see also \citealt{jahnke11}]{peng07}, who have argued that
the local host galaxy - black hole mass correlations can naturally
arise through galaxy mergers without co-evolution between them and no
initial relation. The prediction for that scenario is that the
dispersion of the $M_{\rm BH} - L_{\rm Bulge}$ relation would increase
substantially with redshift, possibly consistent with the large spread
observed in Figure \ref{fg:mbh_msph}. Hot DOGs and QSOs could then be
regarded as the tail of that distribution, namely exceedingly massive
black holes in faint host galaxies.

\subsection{$M_{\rm BH}$ Estimates based on the Local $M_{\rm BH}-M_{\rm Sph}$ Relation}\label{ssec:mbh_case2}

\begin{figure}
  \begin{center}
    \plotone{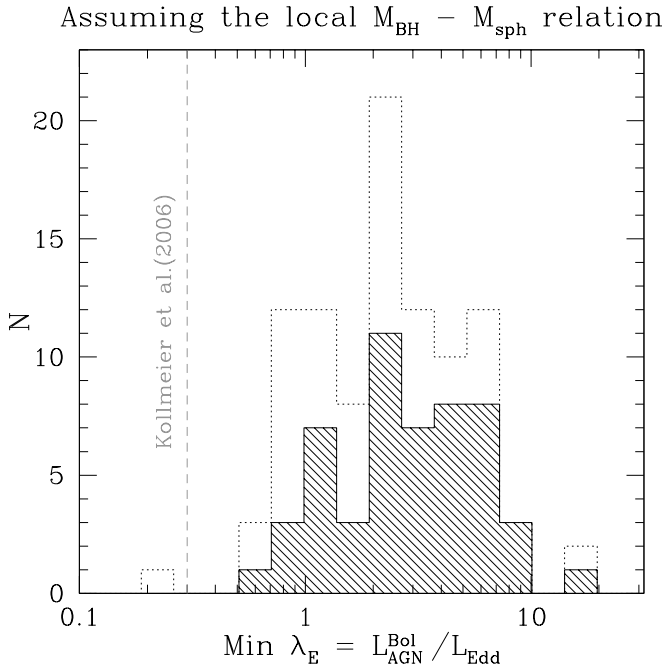}
    \caption{Distribution of the minimum Eddington ratio for the AGN
      in Hot DOGs, assuming that the local relation between $M_{\rm
        Bulge}$ and $M_{\rm BH}$ holds at high redshift. The solid
      hashed histogram shows the distribution of the core sample
      (i.e., W4$<$7.2), while the dashed open one shows the
      distribution of the whole sample. The dashed gray line shows the
      the peak Eddington ratio found by \citet{kollmeier06} for $z\sim
      2$ Type 1 QSOs.}
    \label{fg:min_lrat}
  \end{center}
\end{figure}

In the second scenario, we assume that the local relationship between
the spheroidal component stellar mass and $M_{\rm BH}$ is appropriate
for Hot DOGs (see \S\ref{sec:sed_fits}), and obtain an upper bound of
the SMBH mass from the estimates of the maximal stellar mass. Combined
with the AGN bolometric luminosity of Hot DOGs, this provides an
estimate of the ``minimum'' Eddington ratio ($\lambda_E$) at which the
AGN is radiating. Figure \ref{fg:min_lrat} shows the distribution of
the implied Eddington ratio of Hot DOGs. The peak is about 2--3 times
the Eddington limit, and almost all objects are accreting at
$\lambda_E \geq 1$. For comparison, we show the typical Eddington
ratio of $\sim 0.30$ for QSOs at a redshift of $\sim 2$ determined by
\citet{kollmeier06}.

Although the scenario discussed in \S\ref{ssec:mbh_case1} may seem
more likely, Hot DOGs might indeed be radiating at several times their
Eddington limit. Note the Eddington limit is not necessarily a hard,
physical boundary in this case, since the accretion is most likely not
isotropic. Due to photon pressure, it is possible that the AGN would
be in the process of mechanically expelling material from the regions
close to the central SMBH and, possibly, from the galaxy's potential
well. This would be consistent with the high fraction of extended
Ly$\alpha$ emission reported by \citet{bridge13}, and may relate
specifically to the key stage in the standard galaxy evolution
paradigm where star-formation is being shut down, and thus
transitioning to the QSO stage where it will remain until the
accreting gas supply stops \citep[see, e.g.,][]{hopkins08}. Were we to
assume the evolving $M_{\rm BH}/M_{\rm Sph}$ relation of \citet[][see
  \S\ref{ssec:mbh_case1}]{bennert11b}, our SMBH masses would be
underestimated by a factor of 3.5--5 and the AGN in the Hot DOG
population would still be radiating at, or close to the Eddington
limit, well above the mean QSO Eddington ratio found by \citet[][see
  Fig. \ref{fg:min_lrat}]{kollmeier06}. A correction factor closer to
10 is needed to yield consistency with the mean $\lambda_E$ of
\citet{kollmeier06}. In \S\ref{sec:qso_hd_comp} we discuss the number
density of these objects and further argue that Hot DOGs might indeed
be radiating at high Eddington ratios.

The presence of extended Ly$\alpha$ emission may put interesting
constraints on the lifetime of Hot DOGs if we assume it comes from
winds launched by a super-Eddington accreting AGN. \citet{bridge13}
find the emitting gas has projected velocities of up to several
thousand $\rm km~\rm s^{-1}$. Assuming a constant velocity of $\sim
10,000~\rm km~\rm s^{-1}$, for the gas to have been launched from the
central regions of the galaxy, this stage would have to last
$\gtrsim10~\rm Myr$ to reach a distance of $\sim 100~\rm kpc$. This is
consistent with the lifetimes of $1~\rm Myr < t_{\rm QSO} < 20~\rm
Myr$ estimated by \citet{trainor13} for a sample of hyper-luminous
QSOs (which they defined as $L_{\rm UV}\sim 10^{14}~L_{\odot}$). On
the other hand, if the gas were simply responding to the emission of
the AGN but was not associated to feedback, the light travel time puts
a weak constraint on the lifetime of this phase to $\gtrsim 3\times
10^5~\rm yr$.

\section{Number Densities of Hot DOGs and QSOs}\label{sec:qso_hd_comp}

Several lines of evidence imply that Hot DOGs are indeed powered by
heavily obscured AGN \citep{wu12,eisenhardt12,eisenhardt13}, and so it
is natural to compare their properties to those of comparably luminous
but unobscured Type 1 QSOs, which are much better studied. The
simplest comparison is of their space densities. For this, we focus on
the core sample of 252 W12drops described in \S\ref{ssec:wise_obs}
(i.e., W4$<$7.2). We know that of the 103 core W12drops with {\it{Warm
    Spitzer}}/IRAC observations, 58 have reliable spectroscopic
redshifts. Of these, 52 (90\%) are at $z\geq 1$ and can hence be
considered Hot DOGs, while 43 (74\%) are at $z\geq 2$. Assuming that
objects without spectroscopic redshift measurements have a redshift
distribution approximately equal to that of the spectroscopic sample
\citep{eisenhardt13}, there are 187 Hot DOGs with $z\geq 2$ and
W4$<$7.2. Of them, Hot DOG W2246--0526 ($z=4.6$) is the only one at
$z>4$ and it seems to be somewhat of an outlier based on its
photometric properties, so we will limit the study to the Hot DOGs in
the redshift range $2<z<4$.

We study the effects of the selection function using the 42 core
sample Hot DOGs with reliable spectroscopic redshifts in the range
$2<z<4$ and IRAC observations. Because the Hot DOG selection is based
upon the observed WISE magnitudes and colors, objects identified as
core sample Hot DOGs may not have been recognized as such if located
at a different redshift. This causes a significant sample
incompleteness which we take into account by using the $V/V_{\rm Max}$
method of \citet{schmidt68}. In short, we use the best-fit SED model
of each individual source to evaluate the redshift range, and hence
volume $V$, for which the object could have been detected and
identified by WISE as a Hot DOG within the full volume $V_{\rm Max}$,
corresponding to the redshift range $2<z<4$. We then simply assume
that the intrinsic distribution of sources is uniform across the given
volume $V_{\rm Max}$ such that the effective number surface density of
Hot DOGs is then
\begin{equation}\label{eq:n_hd}
  N_{\rm HD}\ =\ \frac{1}{f_z A_{\rm
      Sky}}\ \sum_{i}\ \frac{1}{V_i/V_{\rm Max}},
\end{equation}
\noindent where $f_z=58/252 = 0.23$ is the fraction of core sample Hot Dogs
with IRAC observations and a reliable redshift measurement, and
$A_{\rm Sky} = 32,000~\rm deg^2$ is the area of the sky surveyed for
Hot DOGs. We estimate the error as
\begin{equation}
  \delta N_{\rm HD}\ =\ \frac{1}{f_z A_{\rm
      Sky}}\ \left(\sum_{i}\ \frac{1}{(V_i/V_{\rm
      Max})^2}\right)^{1/2}.
\end{equation}
Two caveats to this process should be noted. First, we apply the
corrections of \citet[][and Lake, S. private comm.]{lake13} to the
model fluxes before evaluating the selection function to account for
the flux bias of the WISE All-Sky release (see \S\ref{ssec:wise_obs}
for details). And, second, we apply a small correction to the W3 and
W4 model fluxes to perfectly match the observed values at the object's
redshift. Even after applying the flux bias correction, we find that
10 of the 52 Hot DOGs in the core sample with IRAC observations and
spectroscopic $z\geq 1$ (6 of which are in the redshift range $2<z<4$)
were only selected as such because of the stochastic nature of the
flux bias. The SED modeling of these objects, driven primarily by the
IRAC fluxes, predicts W1 and W2 fluxes too bright to be selected as
Hot DOGs. We do not use these objects to estimate $N_{\rm HD}$. Note
the change to $f_z$ from eliminating these 7 objects from the sample
is negligible, decreasing from 0.23 to 0.21.

\begin{figure}
  \begin{center}
    \plotone{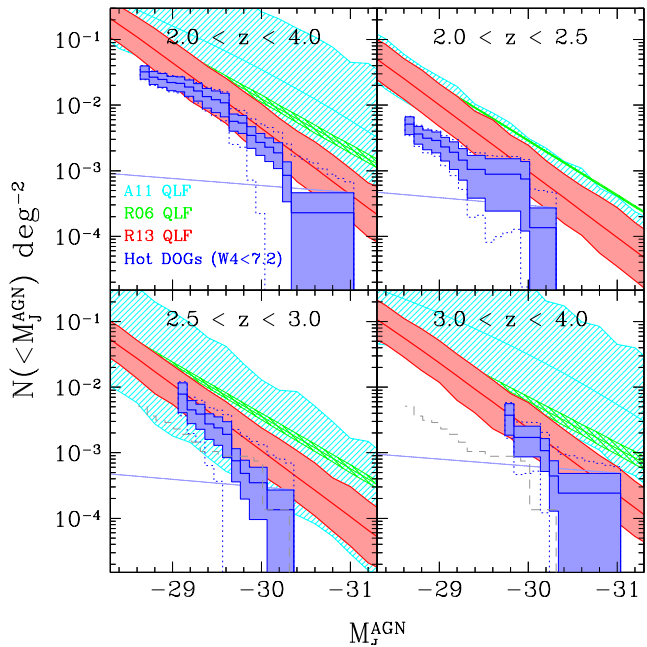}
    \caption{The top left panel shows the number density of Hot DOGs
      in the redshift range $2<z<4$ ({\it{solid blue line}}) corrected
      for volume incompleteness through the $V/V_{\rm Max}$ approach
      of \citet[][see \S\ref{sec:qso_hd_comp} for
        details]{schmidt68}. We also show the region marked by the
      error-bars ({\it{solid blue region}}). The dotted blue lines
      show the uncertainty region taking into account a nominal
      scatter in the AGN covering fractions, as discussed in the
      text. For comparison, we also show the expected number density
      of Type 1 QSOs as predicted by the QLFs of
      \citet[][{\it{green}}]{richards06},
      \citet[][{\it{cyan}}]{assef11} and
      \citet[][{\it{red}}]{ross13}. The remaining three panels show
      the same but for three redshift slices within the full
      range. For comparison, the Hot DOG distribution of the $2<z<2.5$
      slice is shown in the other two slices ({\it{dashed gray
          line}}).}
    \label{fg:n_hd}
  \end{center}
\end{figure}

Using equation (\ref{eq:n_hd}) we find then that the effective number
surface density of Hot DOGs is $N_{\rm HD}=$ 0.032 $\pm $ 0.004 $~\rm
deg^{-2}$, or approximately one for every 31 $\pm $ 4 $~\rm
deg^2$. Figure \ref{fg:n_hd} shows the cumulative, volume-corrected
surface density of Hot DOGs as a function of increasing luminosity of
the obscured AGN component. We parametrize the AGN component
luminosity by $M_J^{\rm AGN}$, the absolute $J$-band magnitude it
would have under no obscuration. A drawback of our $V/V_{\rm Max}$
approach to correct for incompleteness is that it disregards evolution
in the Hot DOG number density between $2<z<4$. Furthermore, using such
a large redshift range can lead to very large correction factors,
which could be introducing significant noise to our estimates. To
avoid these issues, Figure \ref{fg:n_hd} also shows the effective
number of Hot DOGs as a function of the AGN component luminosity for
three redshift slices: $2<z<2.5$, $2.5<z<3$ and $3<z<4$. Due to the
low number of sources, the uncertainties are large in the last
redshift bin. Interestingly, little evolution is observed between the
first two redshifts slices, while some evolution may be present when
compared to the $3<z<4$ slice. The statistics are currently
insufficient to be more definitive about this.

The estimates of $M_J^{\rm AGN}$ from our SED modeling are based on
the assumption of a single AGN covering fraction ($c_f$) for all Hot
DOGs, as discussed in \S\ref{sec:sed_fits}. It is possible, however,
that there is a significant scatter on the $c_f$ of Hot DOGs. To
explore this we re-estimate the uncertainty regions now including the
effects of a scatter in $c_f$ equal to that found by
\citet{roseboom13} for Type 1 QSOs. The scatter in $c_f$ found by
\citet{roseboom13} can be expressed as an uncertainty of 0.5~mag in
the magnitude of any band dominated by the accretion disk emission for
a fixed mid-IR flux. Although the rest-frame $J$-band luminosity is
only partly powered by the accretion disk, we consider the full
0.5~mag uncertainty in our estimates of $M_J^{\rm AGN}$ for
simplicity. Figure \ref{fg:n_hd} shows the uncertainty region
considering the effects of such a scatter. Including this effect
increases the size of the uncertainty region, particularly at the
bright end, but does not qualitatively change our results.

To obtain the number density of comparably luminous Type 1 AGN we use
the functional form of the optically selected QLF of
\citet{richards06} and \citet{ross13}, derived from SDSS observations,
and of the mid-IR and X-ray selected QLF of \citet{assef11}, derived
from the deeper but smaller area observations of the NDWFS Bo\"otes
field.  The brighter end of the luminosity function is much better
constrained by the SDSS observations used by \citet{richards06} and
\citet{ross13} than by the much smaller Bo\"otes field used by
\citet{assef11}, suggesting they are a better comparison to our HyLIRG
sample. However, the mid-IR nature of our Hot DOG selection may be
better represented by the selection function of \citet{assef11}. We
note that \citet{richards06} assumed a flattening of the QLF at
$z>2.4$, motivated by the flatter $4<z<5$ QLF of \citet{fan01}, which
has since been shown to be incorrect \citep{mcgreer13} and an artifact
of a rapidly evolving break luminosity \citep[see
  also][]{assef11}. The QLF of \citet{ross13} is only defined over the
redshift range $2.2<z<3.5$, so we extrapolate the parameters using
their functional form to the modestly larger $2<z<4$ range of Hot
DOGs.

Let $\Phi(M_{J},z)$ be the space density of QSOs at redshift $z$ with
absolute magnitude $M_{J}$. The number surface density of QSOs in the
sky that are brighter than a certain luminosity in the redshift range
$2<z<4$ is then given by
\begin{equation}
  N_{\rm QSO}\ =\ \int_{2}^{4} \int_{-\infty}^{M_{J}}
  \phi(M_{J},z)\ \frac{dV_C}{dz}\ dM_{J} dz,
\end{equation}
\noindent where $V_C$ is the comoving volume. The luminosity functions
of \citet{richards06} and \citet{ross13} are parametrized as a
function of $M_{i'}$, the absolute magnitude in the $i'$-band, which
corresponds to the SDSS $i$-band shifted to $z=2$. We use the
rest-frame AB color of a Type 1 QSO with no host-contamination of
$i'-J=-0.78$ determined by \citet{assef11} to convert between the
absolute $i'$ and $J$-band magnitudes. Figure \ref{fg:n_hd} shows the
cumulative surface density of Type 1 QSOs according to each of the QLF
parametrizations. Each of the uncertainty regions shown in Figure
\ref{fg:n_hd} are estimated by calculating 500 realizations of the QLF
created through re-sampling of each of its parameters according to
their published 1$\sigma$ uncertainties and assuming Gaussian
statistics. The shaded regions show the 68\% confidence interval of
the 500 realizations. Note that as the co-variances between parameters
are neglected, the uncertainty regions could be somewhat
overestimated.

Figure \ref{fg:n_hd} shows that the counts we obtained for Hot DOGs
are quite well matched at the bright end by those predicted by the QLF
of \citet{ross13}, suggesting that Hot DOGs are as common as QSOs of
comparable luminosity. This is roughly consistent with the comparison
to the \citet{assef11} QLF except in the highest redshift range, where
error bars are quite large due to the small size of the NDWFS Bo\"otes
field. Unsurprisingly, there is a significant discrepancy with the
\citet{richards06} QLF, consistent with the large discrepancy between
the latter and the \citet{ross13} QLFs. Also note that the optical
color selection function of \citet{richards06} is least effective at
$z\sim 2.5$ due to confusion with the colors of the stellar locus
\citep{fan99}. The difference in the faint-end of the slopes of the
respective curves at $2.0<z<2.5$ may also suggest that Hot DOGs follow
a different luminosity function than QSOs, although this is not
observed in the higher redshift bins.

Some studies of lower-luminosity QSOs have found that the fraction of
Type 2 AGN decreases strongly as a function of increasing bolometric
luminosity \citep[see, e.g.,][]{ueda03,hasinger04,simpson05,assef13},
implying there should be a very small number of obscured QSOs at the
luminosities of Hot DOGs. This prediction is significantly at odds
with the results shown in Figure \ref{fg:n_hd}, since Hot DOGs appear
to be as common as comparably luminous Type-1 QSOs. This discrepancy
could indicate a reversal in the trends found at lower luminosities,
implying that the fraction of Type-2 QSOs increases with luminosity
towards the upper end of the QLF. Such a reversal has also been
suggested by \citet{banerji12}, although with limited statistics,
based on the obscuration fraction of NIR selected
QSOs. \citet{banerji12} suggests that these reddened, high luminosity
QSOs may probe an evolutionary phase rather than an orientation
effect. Other studies, primarily at lower luminosities, have found
that the obscuration fraction may be a shallow function of the
bolometric luminosity \citep{lusso13} or independent of it
\citep[e.g.,][]{wang06,honig11,lawrence10,lacy13}, so that a high
fraction of obscured QSOs at high luminosities may not be
surprising. The obscuration of the lower-luminosity QSOs is thought to
come from dust primarily in the vicinity of the SMBH, namely the dust
torus, and models that replicate the lowering fraction of obscured
objects with increasing luminosity have been devised for these
structures \citep[see, e.g.,][]{lawrence91,simpson05}. This, however,
may not be the case for Hot DOGs, which may be obscured by a different
dust structure, such as dust on significantly larger physical scales
or with a significantly different geometry or covering fraction,
although we caution the reader that a significantly different covering
fraction/geometry than that implied by our AGN SED template could have
a considerable effect over the estimated $M_J$ value (see also the
discussion in \S\ref{sec:sed_fits}). This could naturally solve the
tension with the obscuration trends found for the lower-luminosity
QSOs, and would imply that Hot DOGs are not the torus-obscured
counterparts of the known Type 1 QSOs of similar
luminosities. \citet{lonsdale14} have come to a similar conclusion for
objects that may be the radio-loud counterparts of Hot DOGs. Measuring
the total dust content from a combination of mid-IR and ALMA sub-mm
observations, \citet{lonsdale14} suggests the torii of these objects
would have to be unrealistically large to explain the high
luminosities observed.

Alternatively, we could consider the possibility discussed in the
previous section that the AGN in Hot DOGs radiate at an Eddington
ratio significantly above unity, in contrast to typical $z\sim 2$ QSOs
which radiate at $\lambda_E\sim 0.30$ \citep{kollmeier06}. We can
speculate then that Hot DOGs are objects going through a phase of
their evolution in which, for a brief period of time, they radiate
well above their Eddington limit. If so, we should compare their
number density to that of lower luminosity QSOs instead. In this case,
we would conclude that Hot DOGs only constitute a small fraction of
SMBHs of moderate to high mass, rather than a large fraction of the
tip of the SMBH mass function. The fraction of QSOs of the same SMBH
mass that are in this ``Hot DOG phase'' would constrain the duration
of the latter relative to the QSO lifetime, and could be discussed in
the context of the timescales estimated at the end of
\S\ref{ssec:mbh_case2}. We, however, refrain from exploring this any
further here since there are too many uncertainties for a meaningful
discussion.

As noted in \S\ref{sssec:opt_spec}, 30\% of objects targeted for
spectroscopy did not yield a successful redshift measurement,
primarily due to optical faintness \citep{eisenhardt13}. This means
that while we considered 58 core sample W12drops with spectroscopic
$z$ and {\it{Warm Spitzer}} imaging, 83 were targeted. Their optical
faintness implies the 25 additional objects have high-z, and here we
have assumed they follow the same redshift distribution of Figure
\ref{fg:zhist}. The effects of a different redshift distribution can
be approximated by modifying $f_z$, but is unlikely to qualitatively
modify our conclusions. For example, if all of these 25 objects had
$2<z<4$ and had the same magnitude distribution, the effects over the
$2 < z < 4$ number density distribution of Figure \ref{fg:n_hd} (top
left panel)can be approximated by making $f_z = ((58+25)/252) \times
(42/(42+25)) = 0.21$ instead of $f_z=0.23$. Alternatively, in the
unlikely scenario that all 25 objects had $z < 2$, we could
approximate the effects by making $fz = 83/252 = 0.33$.

\section{The Environment of Hot DOGs}\label{sec:env_mass}

Recently, \citet{jones14} found evidence of an overdensity of sub-mm
neighbors to a small sample of Hot DOGs using 850$\mu$m observations
with SCUBA-2 at JCMT, and also noted that the overdensity did not show
an angular dependence around the Hot DOGs within 90\arcsec. Assessing
the environments of these objects is, however, not trivial, as their
high redshifts and corresponding faintness makes getting spectroscopic
distances a daunting task. Hence, we must rely on a statistical
approach.

We use the deep {\it{Warm Spitzer}}/IRAC imaging described in
\S\ref{ssec:spitzer_obs} to count the number of galaxies neighboring
Hot DOGs. Of the 96 Hot DOGs (see Table \ref{tab:samples}), we use the
90 that were observed as part of the same snapshot program \citep[ID
  70162; see][for details]{griffith12}. Such imaging is ideal for this
as the peak of the stellar emission is redshifted into the IRAC bands
at the distance of Hot DOGs, allowing us to probe much lower stellar
masses than what, for example, optical imaging would allow us to. We
follow the approach of \citet{wylezalek13}, and use the IRAC imaging
to study the field density in the vicinity of Hot DOGs in comparison
with the field density in two control samples: i) around random
pointings in the {\it{Warm Spitzer}} UKIDSS Ultra Deep Survey (SpUDS,
P.I.:J.~Dunlop) representative of field galaxies, and ii) around
radio-loud AGN in the Clusters Around Radio-Loud AGN survey (CARLA;
\citealt{wylezalek13,wylezalek14}; and see
\citealt{galametz10,galametz13} for spectroscopic confirmation of two
of these clusters). CARLA identifies moderately massive clusters at
high redshift.

\begin{figure}
  \begin{center}
    \plotone{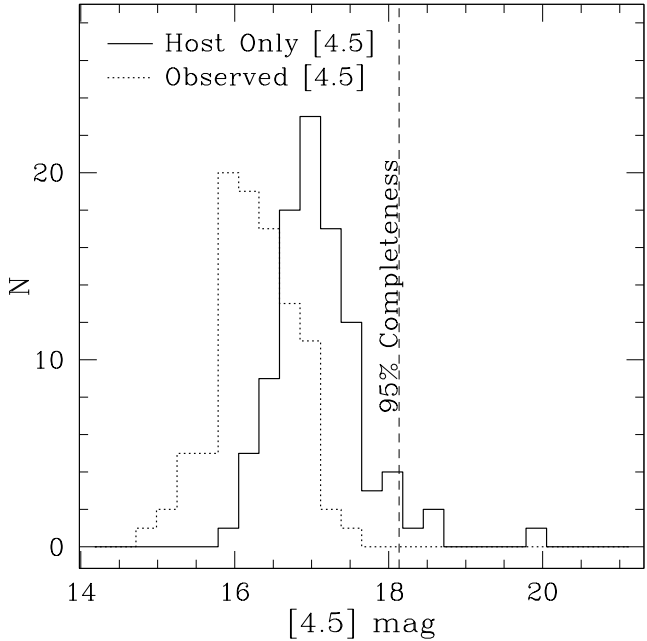}
    \caption{The solid line shows the distribution of the host galaxy
      observed-frame [4.5] magnitudes, obtained from the SED modeling
      described in \S\ref{sec:sed_fits}. For comparison, the dotted
      line shows the observed [4.5] magnitudes, including the
      contribution of the AGN. The vertical dashed line shows the
      $10\mu\rm Jy$ completeness limit of our IRAC imaging.}
    \label{fg:c2_host}
  \end{center}
\end{figure}

We start by counting for each of the 90 Hot DOGs the number of red
galaxies ([3.6]--[4.5]$>$0.37, to select only $z\gtrsim 1$ galaxies)
found in a 1\arcmin\ radius around it. We use a circle with a
1\arcmin\ radius because such size is small enough to fit well within
a {\it{Spitzer}}/IRAC image, but large enough to encompass a typical
mid-IR selected cluster with $\log{(M_{200}/M_{\odot})}\sim 14$ at
$z>1$ \citep{brodwin11,wylezalek13}. We then repeat this around all
420 CARLA radio-loud AGN (RLAGN) and in 437 randomly selected
pointings within the SpUDS survey. We determined that a grid of
19$\times$23$=$437 apertures maximized the number of independent
1$\arcmin$ radii apertures extracted from the SpUDS survey region. The
shallowest IRAC depth of the three samples is that of Hot DOGs, with a
limiting [4.5] flux of 10$\mu\rm Jy$, so we only consider objects down
to that depth in all three samples. We note that the [4.5] magnitudes
of the host galaxies in Hot DOGs, obtained from the SED modeling
described in \S\ref{sec:sed_fits}, are typically $\sim 1~\rm mag$
brighter than the 10$\mu\rm Jy$ depth of our IRAC imaging, as shown in
Figure \ref{fg:c2_host}.

\begin{figure}
  \begin{center}
    \plotone{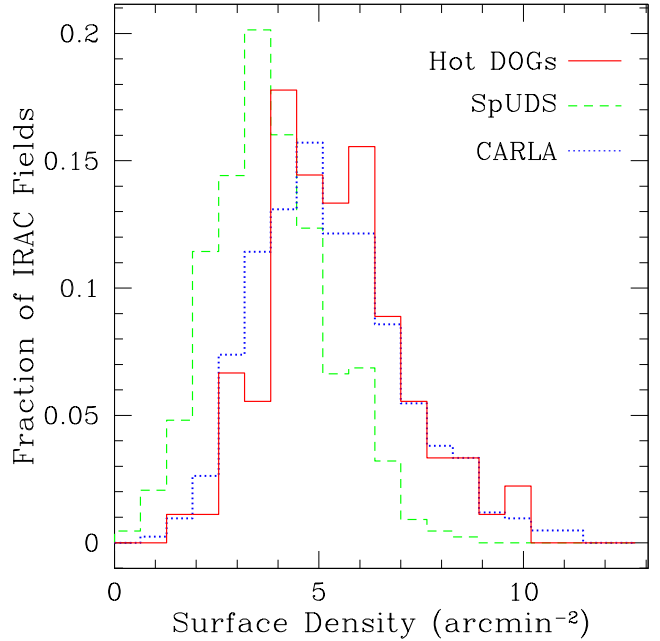}
    \caption{The solid red histogram shows the fraction of Hot DOGs
      with a given number density of objects brighter than 10$\mu$Jy
      at [4.5] and with [3.6]-[4.5]$>$0.37 within 1\arcmin. We only
      consider Hot DOGs with IRAC observations and a reliable
      spectroscopic redshift. For comparison, we also show the
      distribution for all objects in the CARLA survey ({\it{dotted
          blue histogram}}) and for a sample of randomly selected
      positions in the SpUDS survey ({\it{dashed green
          histogram}}). The Hot DOG fields are overly dense compared
      to the random field, and similar to the CARLA fields.}
    \label{fg:hd_irac_number}
  \end{center}
\end{figure}

Figure \ref{fg:hd_irac_number} shows the results of this analysis. The
density of galaxies in the Hot DOG imaging is greater than that in the
SpUDS control sample, suggesting the environment of Hot DOGs is
significantly more dense than that of field galaxies. The Hot DOGs
show good agreement with the CARLA sources, suggesting that Hot DOGs
live in environments similar to those of RLAGN. Formally, a K-S test
shows that the probability of the Hot DOG surface density distribution
being drawn from the same parent population as that of the random
pointings on the SpUDS survey field is $1.2\times 10^{-10}$, while the
probability raises to 0.69 when comparing with CARLA. The agreement
with the CARLA survey fields is unexpectedly good, and leads to the
speculation of whether Hot DOGs could be the precursors of RLAGN. In a
follow-up work we will study in depth the radio properties of Hot
DOGs, and explore this suggestion further \citep{tsai15}.

As discussed earlier, one of the key assumptions of our SED analysis
is that little to no obscuration is present in Hot DOG host
galaxies. Under significant obscuration, Hot DOG stellar masses could
be considerably larger and affect the interpretation of the results
presented in \S\ref{sec:analysis}. We note, however, that the stellar
masses estimated by \citet{debreuck10} for RLAGN in CARLA clusters,
likely the most massive members of the respective clusters, are not
significantly above our upper bound estimates for Hot DOGs. If the
upper bounds were underestimated due to unrecognized stellar
obscuration, Hot DOGs would be expected to live in considerably denser
environments.

To illustrate this point, we note that a study of the luminosity
function of clusters in CARLA, has shown that at redshifts between
$2.6<z<3$, their [4.5] luminosity function is consistent with a
Schechter function with best-fit values of
$\alpha=-1.28^{+0.15}_{-0.10}$ and $m^*_{\rm
  [4.5]}=19.59^{+0.25}_{-0.25}$ \citep{wylezalek14}. Using this
luminosity function, we can estimate a lower bound on the expected
number of galaxies brighter than 10$\mu$Jy in a cluster containing a
given Hot DOG by assuming the Hot DOG is the brightest cluster
galaxy. When we consider all Hot DOGs with IRAC imaging, we find a
marginal agreement between our predicted field densities and observed
density distribution of the Hot DOG fields, with a K-S probability of
0.19 that both are drawn from the same parent population. Yet, if we
assume that fluxes (hence stellar masses) are underestimated by a
factor as small as 1.5, the K-S probability decreases to $6\times
10^{-7}$.

We conclude from these results that while Hot DOGs may live in dense
environments, their field densities are inconsistent with clusters
that host galaxies more massive than the upper bound stellar masses we
estimated for them in \S\ref{sec:sed_fits}. This reinforces our
assumption of little to no host galaxy obscuration and shows that
systematically underestimated stellar masses are not the drivers of
the results discussed in \S\ref{ssec:mbh_case1} and
\S\ref{ssec:mbh_case2}.

\begin{figure}
  \begin{center}
    \plotone{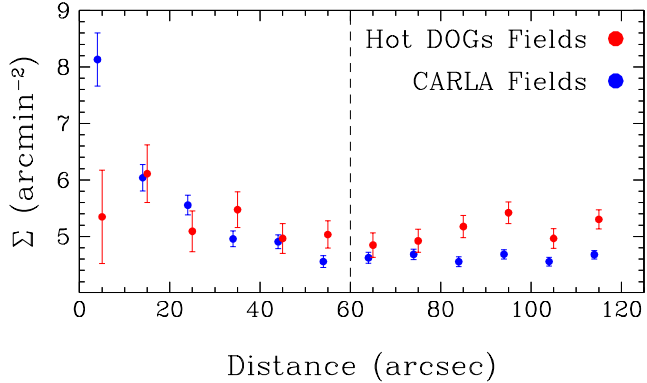}
    \caption{Differential surface density distribution, $\Sigma$, of
      red galaxies surrounding Hot DOGs ({\it{red}}) and CARLA RLAGNs
      ({\it{blue}}) as a function of the distance. The black dashed
      vertical line shows the distance considered for estimating the
      surface densities shown in Figure \ref{fg:hd_irac_number}. The
      points from the CARLA fields have been offset by 1\arcsec\ to
      the left for clarity. Compared to the CARLA fields, the
      environments around Hot DOGs are significantly less
      concentrated, despite the fact that the number density of
      galaxies within 1\arcmin\ are similar. This is consistent with
      the low angular concentration in Hot DOG fields found by
      \citet{jones14} at 850$\mu$m.}
    \label{fg:hd_concentration_angular}
  \end{center}
\end{figure}

Additionally, we can use the IRAC images to look at the concentration
of galaxies in the vicinity of Hot DOGs. To do this we measure the
angular surface density, $\Sigma$, of red galaxies (again selected as
objects with [3.6]--[4.5]$>$0.37 and $f_{\rm{[4.5]}}>10~\mu\rm Jy$),
in the fields of Hot DOGs as a function of the distance to the given
Hot DOG. The results are shown in Figure
\ref{fg:hd_concentration_angular}. For comparison, we repeat the
process in the CARLA fields, centered on the RLAGN targeted by the
survey. We note the redshift distribution of CARLA RLAGN is quite
similar to that of Hot DOGs. Compared to the CARLA fields, the
environments around Hot DOGs are significantly less concentrated,
despite the fact that the number density of galaxies within
1\arcmin\ are similar. Such a difference could imply that Hot DOGs
live in very dense filaments, or possibly in clusters that are in an
earlier state of virialization. Alternatively, such results could be
interpreted as Hot DOGs living in clusters as dense or denser than
those found by CARLA, but that the Hot DOGs are not the central galaxy
of the cluster. This is, however, unlikely, in light of the argument
presented earlier based on the cluster luminosity function. We note
this low angular concentration is consistent with the results of
\citet{jones14} at 850$\mu$m within a similar radial distance of Hot
DOGs. 

\section{Conclusions}

We have presented a detailed study of the observed near-IR through
mid-IR SEDs of a large set of Hot DOGs identified by WISE, focusing on
the subsample with W4$<$7.2 (the core sample). Using the SED templates
of \citet{assef10}, we find that Hot DOGs are generally well fit by a
combination of a luminous and obscured AGN that dominates the emission
at rest-frame wavelengths $\lambda\gtrsim 1\mu\rm m$, and a host
galaxy that dominates the bluer emission. The AGN in Hot DOGs are
among the most luminous AGN known and dominate the bolometric
luminosity of these objects, accounting for $>$97\% of the total
$0.1-30\mu\rm m$ output in all objects.

Using these SED models, we find that the AGN in Hot DOGs display a
large range of obscurations, with $2.5 < E(B-V) < 21.5$, and a mean of
$\langle E(B-V)\rangle =6.8\ (6.4)$ in the core (general)
sample. Using the median dust-to-gas ratio in AGN of
\citet{maiolino01}, $E(B-V)/N_{\rm H} = 1.5\times 10^{-23}~\rm
cm^{2}~\rm mag$, these obscurations correspond to gas column densities
of $1.7\times 10^{23} < N_{\rm H} < 1.4\times 10^{24}~\rm cm^{-2}$, or
over 10 times the $10^{22}~\rm cm^{-2}$ column used to separate Type 1
and Type 2 AGN, reaching into the Compton-thick regime. While
significant host galaxy obscuration is unlikely, such obscuration
would make us underestimate the dust absorption towards the accretion
disk.

We estimate upper bounds on the stellar mass of Hot DOGs using the
rest-frame $K$-band luminosities of the modeled host-galaxy
component. These range from $11 < \log M_{*}/M_{\odot} < 12.5$,
implying Hot DOGs could be some of the most massive galaxies at their
redshifts. It is unlikely these upper bounds are underestimated,
because the environmental densities of Hot DOGs are inconsistent with
those needed to host more massive galaxies at their redshifts (see
\S\ref{sec:env_mass}).

We investigated two approaches to estimate $M_{\rm BH}$ in Hot
DOGs. If we assume the AGN in Hot DOGs radiate at the same $\lambda_E$
of similar redshift QSOs, then Hot DOGs must deviate significantly
from the local $M_{\rm Sph} - M_{\rm BH}$ relation. Such deviations
are also observed for high redshift QSOs, and imply that the SMBH is
assembled considerably before the stellar mass, constraining our
current galaxy evolution models. Alternatively, we can estimate
$M_{\rm BH}$ by assuming Hot DOGs follow the local $M_{\rm Sph} -
M_{\rm BH}$ relation, and derive a minimum $\lambda_E$ for these
objects. If this is the case, AGN in Hot DOGs must be radiating at
significantly super-Eddington ratios. This could imply that Hot DOGs
could be at the transition point where the AGN is possibly expelling
gas from the galaxy and quenching their star-formation, a scenario
that may be consistent with the high fraction of extended Ly$\alpha$
emission found by \citet{bridge13}.

We show in \S\ref{sec:qso_hd_comp} that although very rare, the number
density of Hot DOGs is comparable to that of equally luminous Type 1
AGN in the redshift range $2<z<4$. This suggests that Hot DOGs may not
be the torus-obscured counterpart of the equally luminous Type 1 AGN,
as the dust torus obscuration fraction is expected to be exceedingly
small at these luminosities
\citep[e.g.,][]{lawrence91,simpson05,assef13}. Considering the large
Eddington ratios we estimate for these objects, we speculate that Hot
DOGs may host AGN with significantly less massive SMBHs than the
equally luminous Type 1 AGN, but that they are going through a phase
where their accretion rates are temporarily enhanced, possibly to the
point of quenching star-formation. In this scenario, when radiating at
their ``normal'' Eddington ratios, the AGN in Hot DOGs would
constitute a small fraction of the Type 1 AGN with comparable mass
SMBHs.

Finally, in \S\ref{sec:env_mass} we study the environments of Hot DOGs
using follow-up IRAC imaging. We show that the number of galaxies
within a 1\arcmin\ radius is significantly above the number observed
in random pointings implying Hot DOGs live in dense
environments. Furthermore, we show that the environments are as dense
as those of the clusters identified by the CARLA survey.

Further constraints on the host galaxy properties will allow us to
better place these objects in the galaxy evolution context. The
physical properties we have been able to determine here highlight how
unusual Hot DOGs are among the general galaxy population or even other
previously identified extreme populations (e.g., SMGs), and suggest
that they may represent a pivotal transition in the galaxy evolution
paradigm.

\acknowledgments

We are indebted to all WISE team members. We thank the anonymous
referee for comments and suggestions that helped to improve this
article. RJA was supported by Gemini-CONICYT grant number
32120009. This publication makes use of data products from the
Wide-field Infrared Survey Explorer, which is a joint project of the
University of California, Los Angeles, and the Jet Propulsion
Laboratory/California Institute of Technology, funded by the National
Aeronautics and Space Administration. This work is based in part on
observations made with the {\it{Spitzer Space Telescope}}, which is
operated by the Jet Propulsion Laboratory, California Institute of
Technology under a contract with NASA. Kitt Peak National Observatory
and Cerro Tololo Inter-American Observatory, National Optical
Astronomy Observatory, are operated by the Association of Universities
for Research in Astronomy (AURA) under cooperative agreement with the
National Science Foundation. The WIYN Observatory is a joint facility
of the University of Wisconsin-Madison, Indiana University, Yale
University, and the National Optical Astronomy Observatory. The SOAR
Telescope is a joint project of: Conselho Nacional de Pesquisas
Cient\'ificas e Tecnol\'ogicas CNPq-Brazil, The University of North
Carolina at Chapel Hill, Michigan State University, and the National
Optical Astronomy Observatory. Based partly on observations obtained
at the Hale Telescope, Palomar Observatory as part of a continuing
collaboration between the California Institute of Technology,
NASA/JPL, NOAO, Oxford University, Stony Brook University, and the
National Astronomical Observatories of China.

{\it{Facilities:}} \facility{WISE}, \facility{Spitzer (IRAC)},
\facility{Hale (WIRC)}, \facility{MMT (SWIRC)}, \facility{SOAR
  (OSIRIS)}, \facility{WIYN (WHIRC)}

\end{document}

%% file: tab1.tex
\begin{deluxetable*}{l c c c c c}

  \tablecaption{Sample Definition\label{tab:samples}} 

  \tablehead{
    \colhead{Sample}&
    \colhead{Description}&
    \multicolumn{2}{c}{{\it{All}}}&
    \multicolumn{2}{c}{{\it{IRAC Detected}}}\\
    \colhead{} &
    \colhead{} &
    \colhead{N}&
    \colhead{N$z$}&
    \colhead{N}&
    \colhead{N$z$}
  }

  \tabletypesize{\small}
  \tablewidth{0pt}
  \tablecolumns{6}

  \startdata
  Full Sample W12drop    & Eqn. (\ref{eqn:w12d_sel})    & 934     & 155    & 711     & 115\\
  & & \\
  Full Sample Hot DOG    & $z>1$ and W12drop            & \nodata & 122    & \nodata & \phn96\\
  & & \\
  Core Sample W12drop    & W4$<$7.2 and W12drop         & 252     & \phn95 & 103     & \phn58\\
  & & \\
  Core Sample Hot DOG    & W4$<$7.2 and Hot DOG         & \nodata & \phn77 & \nodata & \phn52\\
  & & \\
  Hot DOGs for number    & $2<z<4$ and                  & \nodata & \nodata & \nodata & \phn42\\
  density analysis (\S\ref{sec:qso_hd_comp}) & Core Sample Hot DOG & \\

  \enddata

  \tablecomments{All results presented in this article are based on
    objects detected by {\it{Warm Spitzer}}/IRAC (see
    \S\ref{ssec:spitzer_obs}). The definition of the Hot DOG
    population presented in this table reflects the definition adopted
    for this work but is somewhat more restrictive than those used
    elsewhere (e.g., as in \citealt{wu12}; see
    \S\ref{sec:analysis}). Note that for samples where $z$ is not
    required for their selection, $N$ is the total number of objects
    regardless of whether they have been observed spectroscopically,
    while $Nz$ shows the number of objects with a measured redshift.}

\end{deluxetable*}

%% file: tab2.tex
\begin{deluxetable}{l c c c c}

  \tablecaption{Ground-Based NIR Follow-Up\label{tab:nir_mags}} 

  \tablehead{
    \colhead{WISE ID}&
    \colhead{Band}&
    \colhead{Mag\tablenotemark{$\dagger$}}&
    \colhead{Unc.}&
    \colhead{Instrument\tablenotemark{$\ddagger$}}
    }

  \tabletypesize{\small}
  \tablewidth{0pt}
  \tablecolumns{6}

  \startdata
  WISEJ000431.34--192301.8       &$H$   &    19.453 &  0.125 & A    \\
  WISEJ000709.03+730831.2        &$J$   &    21.795 &  0.630 & B    \\
  WISEJ002659.24+201556.2        &$J$   &    20.270 &  0.117 & B    \\
                                 &$Ks$  &    18.833 &  0.142 & C    \\
  WISEJ002933.06+020505.4        &$J$   &    19.794 &  0.166 & B    \\
  WISEJ012611.98--052909.6       &$J$   &    19.861 &  0.135 & B    \\
  WISEJ013400.59--260726.5       &$J$   &    20.163 &  0.123 & A    \\
  WISEJ014747.59--092350.5       &$J$   & $>$23.000 &        & B    \\

  \enddata

  \tablecomments{ Table \ref{tab:nir_mags} is published in its
    entirety in the electronic edition of ApJ. A portion is
    shown here for guidance regarding its form and content.}

  \tablenotetext{$\dagger$}{1$\sigma$ upper bounds shown for undetected
    sources.}  

  \tablenotetext{$\ddagger$}{Telescope and instrument used for the
    observations, defined as {\bf{A}}: SOAR/OSIRIS; {\bf{B}}: WIYN
    3m/WHIRC; {\bf{C}}: Hale 200-inch/WIRC; and {\bf{D}}: MMT/SWIRC.}

\end{deluxetable}